\documentclass{iau} 
\usepackage{graphicx}
\usepackage{caption}
\usepackage{subcaption}

\title[Stellar eclipses at radio]{The influence of eclipses in the stellar radio emission}

\author[C.~L.~Selhorst \& A.~Valio]{Caius~Lucius~Selhorst$^{1,2}$ \and Adriana~Valio$^3$}

\affiliation{$^1$ NAT - N\'ucleo de Astrof\'isica Te\'orica -€" Universidade Cruzeiro do Sul \\ S\~ao Paulo, SP, Brazil \\ email: {caiuslucius@gmail.com}\\[\affilskip]
$^2$IP\&D - Universidade do Vale do Para\'iba - UNIVAP \\ S\~ao Jos\'e dos Campos, SP, Brazil\\
$^3$CRAAM - Universidade Presbiteriana Mackenzie, S\~ao Paulo, SP, Brazil}

\pubyear{2016}
\volume{328}  
\jname{Living Around Active Stars}
\editors{D. Nandi, A. Valio \& P. Petit, eds.}
\begin{document}

\maketitle

\begin{abstract}
Here we simulate the shape of a planetary transit observed at radio wavelengths. The simulations use a light curve of the K4 star HAT-P-11 and its hot Jupiter companion as proxy. From the HAT-P-11 optical light curve, a prominent spot was identified  ($1.10~R_P$ and $0.6~I_{C}$). On the radio regime, the limb brighting of 30\% was simulated by a quadratic function, and the active region was assumed to have the same size of the optical spot. Considering that the planet size is  6.35\% of the the stellar radius, for the quiet star regions the transit depth is smaller than 0.5\%, however, this value can increase to $\sim2\%$ when covering an active region with 5.0 times the quiet star brightness temperature.

\end{abstract}

\firstsection 
\section{Introduction}

Since the discovery of the first exoplanets in the nineties, the number of confirmed ones exceed 3500 at this time (http://exoplanet.eu). Part of this success can be addressed to dedicated projects like  HARPS and Kepler. 

Some of these exoplanets can observed by  the dimming of the light from the parent star during the planetary transit (e.g., \cite[Alonso \textit{et~al.} 2004]{Alonso2004}). Besides planet detection, the transit can be used to detect spots on the stellar surface (\cite[Silva 2003]{Adriana2003}) and estimate the stelar activity (\cite[Silva-Valio {\it et~al.} 2010]{Adriana2010}). 

Despite the great number of exoplanets, the observations are still restricted to the optical wavelengths range.  Although  recently, a transit observation was reported in X-ray (\cite[Poppenhaeger {\it et al.} 2013]{Poppenhaeger2013}), with interesting results indicating that the hot Jupiters  atmosphere can be broader in X-ray than the observed in the optical.   

In \cite[Selhorst {\it et al.} 2013]{Selhorst2013}, the authors considered the physical contributions of the planetary transits observations at radio frequencies. However, the attempts to detect the exoplanet radio emission were restricted to trying to observe the emission from the planet atmosphere, but, without success (e.g., \cite[Hallinan {\it et al.} 2013]{Hallinan2013}).

Here, we present a simple model to estimate the influence of eclipses in the stellar radio emission based on the observed optical light curves. 

\section{Simulations}

To model the optical limb darkening observed in the stellar light curves, \cite{Adriana2003} used the following quadratic function: $I(\mu)/I(1)=1-w_1(1-\mu)-w_2(1-\mu)^2$, where $\mu$ is the cosine of the angle between the line of sight and the normal to the local stellar surface. 

\begin{figure}[!h]
  \centering
  \includegraphics[width=12.cm]{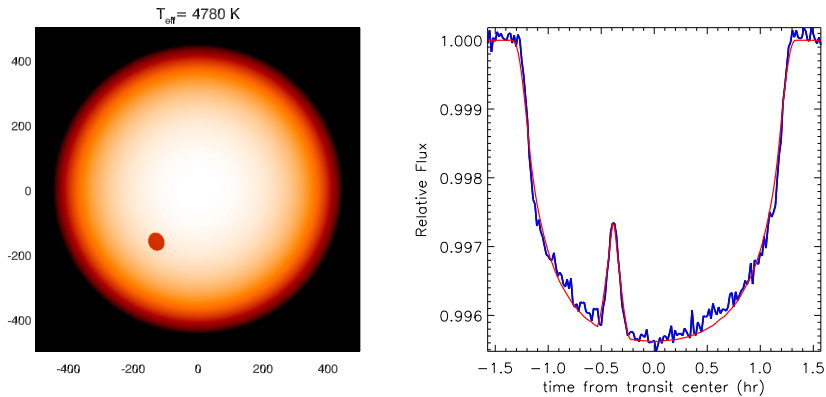} 
   \caption{Left panel: Synthesised model of HAT-P-11 with a dark spot ($0.6~I_{C}$) aiming to reproduce the observed transit light curve. Right panel:  The simulated transit light curve (red) in comparison with the observed one (blue). }
   \label{Fig01}
\end{figure}

In order to test the model, we choose a light curve of the star HAT-P-11 and its planet HAT-P-11b (\cite[B\'eky {\it et al.} 2014]{Beky2014}), which present a single dark spot . The HAT-P-11 is a K4 star, with a temperature of $4780.0~(\pm~50.0)~K$ and a radius of $0.75~(\pm0.02)~R_\odot$. Its planet is a hot Jupiter with a semi-major axis of $\sim 0.05~AU$ and a radius of $R_P\sim0.42~R_J$ or 6.35\% of the the stellar radius. The left panel of Figure~\ref{Fig01} shows the synthesised model (\cite{Adriana2003}) of HAT-P-11 with a dark spot aiming to reproduce the observed transit light curve, whereas the right panel compares the simulated transit light curve (red) with the observed one (blue). The stelar limb darkening was simulated with $w_1=0.3$ and $w_2=1.0$, whereas, the spot had $1.10~R_P$ and $0.6~I_{C}$, where  $R_P$ is the planet radius and $I_C$ is the stellar central intensity.

While the stelar optical observations show characteristic limb darkening, solar maps at radio wavelengths present a limb brightening  (see \cite[Selhorst {\it et al.} 2003]{Selhorst2003} and references therein). A simple change in the equation operators can reproduce the observed limb brightening, i.e., $I(\mu)/I(1)=1+w_1(1-\mu)+w_2(1-\mu)^2$. Setting $w_1=0.1$ and $w_2=0.2$ the synthesised model present a limb brightening 30\% greater than the central temperature, that is compatible with the solar radio observations (\cite{Selhorst2003}).

For the radio simulations, we assume an opaque planet with the same radius of the optical observations, however, the X-ray observations ({Poppenhaeger2013}) suggested that the atmosphere of hot Jupiters can be optically thick at radio. We also adopt the active region with the same size of the spot simulation above and its temperature was defined as uniform. 

\begin{figure}[!h]
  \centering
  \includegraphics[width=12.cm]{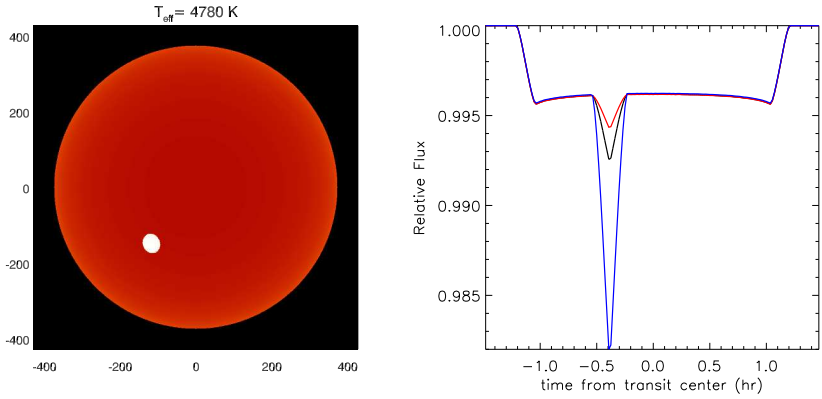} 
   \caption{Left panel: Synthesised model of HAT-P-11 free-free radio emission. Right panel: light curves considering active regions temperatures of 1.5, 2.0 and 5.0 times the quiet star temperature, which can be observed at the Sun at mm and cm wavelengths. 
}
   \label{Fig02}
\end{figure}

The left panel of Figure~\ref{Fig02} shows a synthesised image of the star with limb brightening and the active region of the same size and position as simulated in the optical. The active region brightness temperature was chosen based on the solar observations, where active regions with maximum brightness due to free-free emission varies from 0.2 up to 5.0 times the quiet values depending on the wavelength (e.g., \cite[Silva 2005]{Adriana2005}, \cite[Selhorst {\it et al.} 2008]{Selhorst2008}). The effects in the light curves caused by active regions with  temperatures of 1.5, 2.0 and 5.0 times the quiet star temperature are plotted on the right panel Figure~\ref{Fig02}.

\section{Discussion and Conclusions}

Despite the great number of planetary systems discovery in the last two decades, no detection yet has been reported at radio wavelengths. However, new radio interferometers, like ALMA and the future SKA, may be able to fill this gap.  In this work we simulated the use of planetary transits to investigate the radio emission coming from active regions. 

To test the simulations, the Silva (2003) spot model was applied to determine parameters of the spot observed in the light curve of HAT-P-11. The simulations suggested that the starspot is bigger than the planet size ($1.10~R_P$) and cold with only  $0.6~I_{C}$. This low temperature could be due to much more intense magnetic field than those observed in sunspots.

The radio simulations were performed with a quadratic function to estimate the limb brightening, while the active region was assumed to have an uniform temperature. The reduction in the light curve during the transit was smaller than 0.5\% in the quiet star region, however the depth increased to ~2\% when the planet crossed the active region with a temperature 5.0 times the quiet star temperature. These values are consistent with the values suggested in the previous work (\cite{Selhorst2013}) that used 17~GHz solar maps as a proxy for  stellar emission.

\begin{acknowledgements}
C.L.S. acknowledge financial support from the S\~ao Paulo Research Foundation (FAPESP), grant \#2014/10489-0. 
\end{acknowledgements}

\end{document}